\documentclass[12pt]{iopart}
\usepackage{amsfonts, amsmath, amssymb, amsthm, mathtools}
\usepackage{graphicx}
\usepackage{natbib}
\usepackage{algorithm, algorithmic}
\newtheorem{example}{Example}[section]

\begin{document}

\title[Kernel Code for DNA Digital Data Storage]{Kernel Code for DNA Digital Data Storage}
\author{NallappaBhavithran G$^1$ and Selvakumar R$^{2, *}$}

\address{1 Vellore Institute of Technology, Vellore, India. Email : nallabhavitran@yahoo.com}

\address{2 Vellore Institute of Technology, Vellore, India. Email : rselvakumar@vit.ac.in}

\vspace{10pt}
\address{* Corresponding author}

\begin{abstract}
The biggest challenge when using DNA as a storage medium is maintaining its stability. The relative occurrence of Guanine (G) and Cytosine (C) is essential for the longevity of DNA. In addition to that, reverse complementary base pairs should not be present in the code. These challenges are overcome by a proper choice of group homomorphisms. Algorithms for storage and retrieval of information in DNA stings are written by using kernel code. Complexities of these algorithms are less compared to the existing algorithms. Construction procedures followed in this paper are capable of constructing codes of required sizes and Reverse complement distance.
\end{abstract}

%
\vspace{2pc}

\section{Introduction} 
A way of storing information in DNA was first proposed by (\cite{bib1}). He exploited the human genome to hide information, since it is hard to identify the gene to crack. This paved a path for biodegradable DNA-based storage technologies to emerge. It can be utilised to resolve the problems associated with silica-based systems. Moreover, DNA information is long-lasting since humans can recover information from fossils after millions of years. When it comes to the storage of information, cost and error correction techniques are always a factor to consider. Even though DNA synthesis costs are now high, it is predicted to be dropped in 5 years, according to Molecular Information storage programs(\cite{bib2}). In (\cite{bi3}; \cite{bi4}), a thorough analysis of DNA as a storage system has been conducted, and the advantages and disadvantages of the system are discussed.\\
DNA is in a helical structure containing four nucleotide bases, A-Adenine, G-Guanine, C-Cytosine, and T-Thymine, called elementary alphabets, which are held by an alternating strand of sugar and phosphate. The helical structure consists of two long strands of DNA bases held together by the Watson-Crick complementary bases A = T(two hydrogen bonds) and G $\equiv$ C(Three hydrogen bonds) denoted as A$^c$ = T, G$^c$ = C and vice versa in this paper. A short single strand of DNA, synthesized using DNA syntheses, is called oligos. Here, elementary alphabets {A, G, T, C} are formed in a growing chain of nucleotides by passing one at a time with phosphoramidite as its backbone. The resulting full-length oligos are stored in the living e-coli bacteria for later use. The oligo is being created in between the range of 100 - 1000 base pairs(bps) because of the high synthesizing cost and, also synthesizing error rates depending upon the length of the oligos. In DNA storage, information is mapped to base pairs by encoding techniques(\cite{bib3}) and oligomers (also known as primers) are created based on those base pair blocks. For retrieval, the e.coli is split into tiny fragments, and the base pairs on such pieces are read via sequencing, and the results are combined to provide the information (\cite{bib4}; \cite{bib5}). Based on the synthesis and storage process, we have to formulate a perfect code to reduce the occurrence of errors.\\
First, the set of all possible information has to be converted into the form of nucleotide bases. Moreover, redundancy bases are added to the converted data to prevent errors and data loss. These data transfers and redundancy are achieved through functions that preserve part of the data's algebraic characteristics. The range of this function is called the code. Many properties are imposed on these functions to enrich the algebraic structure of the code, which in turn increases the stability and reduces the secondary structure formation in DNA. (\cite{bib6}) proposed an encoding technique for creating cyclic codes for all odd-length over the ring $F_4[U]/<U^3>$ . The cyclic nature of the code made it possible to determine whether a codeword in the code has its reverse complement or not. This code satisfies the bounds discussed in (\cite{bib7}; \cite{bib8}; \cite{bib9}). In the paper (\cite{bi10}), the languages that will overcome the limitations of DNA storage are analysed and in (\cite{bi11}), cyclic codes are constructed based on languages.\\
In (\cite{bib10}; \cite{bib11}; \cite{bib12}; \cite{bib13}), many experiments were conducted to store data in DNA ranging from 600 Kb - 200 Mb. In 2016, (\cite{bib14}) proposed an idea which provides random access and rewritability. In the paper(\cite{bib14}), six institutions’ budget information was saved in 27 DNA strands with 1000 bps for each strand, and both ends of the strands is an address of length 20. These addresses are drawn from a set of 50\% GC-content, uncorrelated components with a large hamming distance and a short reverse complement distance. Then the information is encoded using those addresses and combined into a long DNA strand. To eradicate the formation of secondary structure in the strand, the codeword is cyclically rotated with a distance of 3 base pairs. For updating the budget information, OE-PCR and gblock methods are used.\\
In (\cite{bib15}), information is taken from a set of q non-negative integers and sliced this into small pieces of fixed length(\textit{l}). The profile vector is made up of $q^l$ coordinates, each of which reflects the number of unique elements created by various lexicographical combinations. The two-step encoding technique decodes the profile vector from $\hat{p}$(x). The first stage is to create profile vectors, and the second step is to utilise Varshmov’s Decoding method. The main obstacle here is choosing the code wisely, to pick the elements with different profile vectors.
This paper(\cite{bib15}) suggests an alternative encoding technique for (\cite{bib14}).\\
In this paper, we are introducing a new technique which eradicates the formation of secondary structures, also increases the stability of the code by imposing GC-content and reverse complement constraints by proper choice of group homomorphisms. In section \ref{sec:Types of errors}, we discuss the problems which arise while using DNA as a storage system.  In section \ref{sec:Constrains}, we analyse those problems discussed in section \ref{sec:Types of errors} by imposing constraints. We introduce the kernel code in section \ref{sec:Kernel Code}, which is used in section \ref{sec:construction} for the construction of the algorithm. Finally, in section \ref{sec:construction}, we give our constructed algorithm and a comparative analysis of our technique with the existing ones.

\section{Types of Errors}\label{sec:Types of errors}
\subsection{Synthesizing Errors}\label{subsec: Synthesizing Errors}
Since the primer is created artificially, the occurrence of errors is unavoidable; the basic errors are:
\begin{itemize}
    \item Insertion: occurs when an additional nucleotide is inserted into the primer.
    \item Deletion: occurs when a nucleotide is deleted from the primer.
    \item Substitution: occurs when a nucleotide is replaced wrongly with another nucleotide in the primer.
\end{itemize}
Even though the insertions and deletions affect the length of the primer, in the long run, these errors co-occur. So, it is hard to identify the place of the error occurrence. To rectify these errors, one should compare similarities between two DNA strings. For this, Levenshtein distance is used. This distance specifies the minimum number of steps required to match one string to another, where each step is an insertion, deletion or substitution. 
Example: d$_l$(GATCTG, GTCAG) = 2. Here, to match the two codewords, the second letter has to be inserted/deleted based on the requirement, and 5$^{th}$ word has to be substituted from ‘T’ to ‘A’. Since calculating Levenshtein distance is an NP-hard problem, approaches based on hamming distance(section 3.1) and uncorrelation are used mostly.
\subsection{Sequencing Errors}\label{Sequencing Errors}
Tantum errors are a type of sequencing error that occurs when reconstructing a whole sequence from fragments. The error is caused when a single base pair or a substring of the primer is replicated and attached adjacent to itself. For example, A tantum error of length 2 occurs if  \textbf{AG}ACAGTG is changed to \textbf{AGAG}ACAGTG. Here, note that AG also appeared in the 7th and 8th places of the sequence(AGAGACAGTG). But, it is not considered a tantum error since it’s not occurring consecutively. Even though this error rectification is mainly done for treatments and some other medical purposes, one has to consider this error for the prolonged durability of DNA-based storage systems(\cite{bib16}).\\ 
There are some more errors which occur while sequencing. One such error is the wrong or unwanted hybridisation due to the influence of some external factors. This results in the formation of secondary structures (folds) in the long strand of bps. The presence of the complementary bases of the prefix (or suffix) of a DNA strand in the same strand leads to folds. Since small strand primers can be sequenced to generate a long stable strand, complement substrings should not be included during the concatenation of two strings.\\
i.e., Let X, Y, Z $\in$ $\mathcal{C}$, where $\mathcal{C}$ is a DNA code of length n $\in$ $\mathbb{N}$ and $\theta$ be an automorphism on $\sum_{DNA}^n = \{A, G, T, C\}^n$ such that
$$ \theta(a) = \text{Reverse complement of a, for every } a \in \sum_{DNA}^n$$\\
Then, $\theta(X)$ should not be a substring of ZY. Several constraints stated in section 3 are implemented on DNA code words to rectify these errors.
\section{Definition And Constraints DNA codes}\label{sec:Constrains}
 A \textit{DNA code} $\mathcal{C}_{DNA}(n, M, d)$  $\subset$ $\sum_{DNA}^n$ = $\{A, T, G,$ $C\}^n$(nucleotide bases) where each DNA codeword is of length n, size M and minimum distance d. The \textit{Hamming distance} H($x,y$) between two codewords is the number of distinct elements in those two codewords.The \textit{reverse} of a codeword $x = (x_1, x_2, \dots, x_n)$ is $(x_n, x_{n-1}, \dots, x_2, x_1)$ and denoted by $x^R$. Similarly, the \textit{reverse complement}  of a codeword $x = (x_1, x_2, \dots, x_n)$ is $(x_n^c, x_{n-1}^c, \dots, x_2^c, x_1^c)$ and denoted by $x^{RC}$.There are four constraints one should see while constructing a code.They are as follows:
\subsection{Hamming Distance constraint} \label{Hamming Distance constrait}
Let $\mathcal{C}$ be a \textit{DNA code}, $H(x, y) \geq d,\text{  } \forall x,y \in \mathcal{C}$ with $x\neq y$. Then we call this distance d the minimum hamming distance of code $\mathcal{C}$. This constraint is inherited from coding theory and helps determine the number of errors this code can correct.
\subsection{Reverse constraint}\label{Reversecon}
 Let $\mathcal{C}$ be a \textit{DNA code}, $H(x^R, y) \geq d,\text{  } \forall x,y \in \mathcal{C}$. This constraint paved the way for the construction of reverse complement constraint. One can use the bound of this constraint and implement it on the other. 
 \subsection{Reverse Complement constraint}\label{Reversecomple}
 Let $\mathcal{C}$ be a \textit{DNA code}, $H(x^{RC}, y) \geq d,\text{  } \forall x,y \in \mathcal{C}$. This constraint helps to reduce unwanted hybridization errors discussed in section \ref{Sequencing Errors}.
 \subsection{GC-content constraint}\label{GCcon}
 Let $x \in \mathcal{C}$, where $\mathcal{C}$ is a DNA code. The total amount of G and C in x is called the GC-weight ($w_{GC}$(x)) of the codeword. And, the GC content of the codeword is the percentage of $\frac{GC-weight}{\text{length of the codeword}}$.  We say $\mathcal{C}$ satisfies the GC-content constraint, only if the GC-content of every codeword in $\mathcal{C}$ is equal to a fixed number of w.\\ 
 Example: Let AAGCT $\in$ $\mathcal{C}$, then  $w_{GC}$(AA\textbf{GC}T ) = 2 and GC-content = $\frac{2}{5}$ × 100\% = 40\%. 

\subsection{Correlation constraint}\label{correl}
Let $\mathcal{C}$ be a DNA code of length n and x, y $\in$ $\mathcal{C}$. The correlation of x and y is denoted as x $\circ$ y, and defined as follows:
\begin{equation}
 (x \circ y)[i] = \left\{\begin{array}{lll}
                1 & \textnormal{if} & x[i:n] = y[1:n-i] \\
               0 & \textnormal{} & \textnormal{otherwise}
            \end{array}\right.
\end{equation}
(i.e), Example: Let $X = CATCGT, \quad Y = TCGTAC,$ $ X, Y \in \mathcal{C} , \text{then } X \circ Y = 001001$.
\[
 \begin{array}{cccccccccccc}
      X=C&A&T&C&G&T&&&&&&\\
      Y=T&C&G&T&A&C&&&&&&0\\
      &T&C&G&T&A&C&&&&&0\\
      &&T&C&G&T&A&C&&&&1\\
      &&&T&C&G&T&A&C&&&0\\
      &&&&T&C&G&T&A&C&&0\\
      &&&&&T&C&G&T&A&C&1\\
 \end{array}
\]
These constraints are being implemented in our DNA code by using the group homomorphism and the error-correcting properties of the kernel codes (\cite{bib17}). 
\section{Kernel Code}\label{sec:Kernel Code}
\textbf{Definition 1}: Let $\mathcal{G} = \mathcal{G}_1 \times \mathcal{G}_2 \times \cdots \times \mathcal{G}_n$, where $\mathcal{G}_i's$ are groups and ($\mathcal{S}, *$) be an abelian group with identity element \textit{e}. Then $\mu : \mathcal{G} \rightarrow \mathcal{S}$ such that $\mu(g_1, g_2, \dots, g_k) = \mu_1(g_1)*\mu_2(g_2)* \dots *\mu_n(g_n)$, where $g_i \in \mathcal{G}_i$ and $ \mu_i: \mathcal{G}_i \rightarrow S$ is a homomorphism for all $ i = 1\text{ }to\text{ }n$. This construction states that $\mu$ is a homomorphism. The kernel of the homomorphism $K = \{g \in \mathcal{G}/\mu(g) = e\} $ is called the Kernel Code(\cite{bib17}).\\
Example: Let us consider ($\mathcal{Z}_2, \cdot_2$), ($\mathcal{Z}_4, \cdot_2$), ($\mathcal{Z}_2, \cdot_2$), ($\mathcal{Z}_2, \cdot_2$) as $\mathcal{G}_1$, $\mathcal{G}_2$, $\mathcal{G}_3$, $\mathcal{S}$ respectively, and $\mu$: $\mathcal{Z}_2 \times \mathcal{Z}_4 \times$ $ \mathcal{Z}_2 \to \mathcal{Z}_2$ such that $\mu_1(0) = 0$, $\mu_1(1) = 1$, $\mu_2(0) = 0$, $\mu_2(1) = 1$, $\mu_2(2) = 0$, $\mu_2(3) = 1$, $\mu_3(0) = 0$, $\mu_3(1) = 1$. Then the kernel Code of this homomorphism is \{000, 020, 110, 130, 101, 121, 011, 031\}.
\subsection{Concatenated Kernel Code}\label{Concatenated Kernel Code}
Concatenated kernel code is a double-layer construction. Its inner layer is the kernel code itself, as mentioned in section \ref{sec:Kernel Code}, with a slight change that all $\mathcal{G}_i's$ should be the same abelian group $\mathcal{G}$ (\cite{bib18}).\\
Its outer layer is the image of $\mu' : \mathcal{G}^k \rightarrow \mathcal{G}^n$, defined as $\mu'$($g_1,g_2,\dots,g_k$) = $(g_1,g_2,\dots,g_k$, $h_1(g_1,g_2,\dots,g_k)$ $,\dots,h_{n-k}(g_1,g_2, \dots,g_k))$ where $\mathcal{G}$ be an abelian group and $h_i'$s are homomorphism defined based on the error correction.\\ 
The images of the kernel code over $\mu'$ are called concatenated kernel code.\\
Example: Let’s take the $\mathcal{G}$ as $\mathcal{Z}_3$, and the information is mapped to $\mathcal{S}$ as $\mathcal{Z}_3$. For the case, k=3 and n=5,  $\mu_i$ be identity homomorphisms (i.e), $\mu_i$(0) = 0, $\mu_i$(1) = 1,$\mu_i$(2)=2 for i=1,2,3 and $h_1(g_1,g_2,g_3)$=$g_1+g_3$ and $h_2(g_1,g_2,g_3)$ = $g_2+g_3$. \\
The kernel Code is {000, 012, 021, 102, 111, 120, 201, 210, 222}.
For, $\mu$(1, 0, 2) = (1, 0, 2, 1+2, 0+2) = (1, 0, 2, 0, 2).\\
The concatenated kernel code is { 00000, 01220, 02110, 10202, 11122, 12012, 20101,21021,22211}.\\
Here, the kernel of $\mu$ is used for creating the Reverse complement distance of size 2. The concatenated mapping $\mu'$ is used for maintaining the GC content of the code.

\section{Construction of DNA codes}\label{sec:construction}
Here information is considered as a block, of length \textit{l}. Each element of the block is from the same finite abelian group $\mathcal{G}$. Corresponding to all possible information, a  DNA code of length n has to be generated to preserve GC-constraint(section \ref{GCcon}), Reverse Complement(section \ref{Reversecon}) and less correlation(section \ref{correl}). To encode the information, kernel codes(section \ref{sec:Kernel Code}) have been used.\\ 
The algorithm goes as follows.
\begin{enumerate}
    \item The information is mapped to a subset of the kernel code(inner code).
    \item Based on the constraints stated in section \ref{sec:Constrains}, homomorphisms are defined over the kernel code(outer code).
    \item Now, the obtained codeword is mapped onto the corresponding base pairs.
\end{enumerate}
\begin{algorithm}
  \caption{Getting required subset of kernel code}\label{alg:Kernel_sub}
  \begin{algorithmic}[1]
     \REQUIRE{Finite Group and $n$} $\quad$ \algorithmiccomment{$n$ is the length of the codeword}
     \ENSURE {Subset of Kernel Codes K}
     \FOR{i = 1 to n + 1}
        \STATE $\mu_i$ = Make the possible Homomorphism.
     \ENDFOR
    \STATE  Compute C the set of all elements in the Cartesian product of $\underbrace{\mathcal{G}\times \mathcal{G} \times \cdots\times \mathcal{G}}_{\text{n times}}$
    \FOR{j = 1 to $\|C\|$}
    \IF{$\mu_1(g_1)\mu_2(g_2)\dots\mu_n(g_n) = 0\text{ } \AND \text{ } \mu_(g_1) = 1$}
    \STATE \textit{Add to set K} 
    \ENDIF
    \ENDFOR
    \RETURN K  \hfill \algorithmiccomment{The required subset of kernel codes}
  \end{algorithmic}
  \end{algorithm}
The construction starts by choosing the required subset of the kernel code to which the information will be encoded. For that, the length of the DNA codeword(say n) is fed to algorithm \ref{alg:Kernel_sub} with the desired group. Note that, n  $\geq$ \textit{l} + 1 where \textit{l} is the length of the information. The output will be the desired subset that starts with 1, and the length of each element in that subset will be n + 1.\\
Example: For n=3, $\mathcal{G} = \mathcal{Z}_2$. The length of the kernel code is n+1 = 4 and 
\begin{equation}\mu: \mathcal{Z}_2 \times \mathcal{Z}_2 \times \mathcal{Z}_2 \to \mathcal{Z}_2.\end{equation}
Kernel code K over $\mu$ = \{0000, 1100, 1010, 1001, 1111, 0110, 0101, 0011\} and the required subset S = \{1001, 1010, 1100, 1111\}.\\
\begin{algorithm}
    \caption{Mapping information \\to the subset of kernel code}\label{alg:Kernel_map}
  \begin{algorithmic}[1]
     \REQUIRE K, a, n  \hfill \algorithmiccomment{//a is the Information Sequence}
     \ENSURE {The map of the word(a) to K}
     \STATE $\text{Map}(a)$
     \STATE $ l \gets length(a)$
     \IF{$l+1 = n$}
     \FOR{element in K}
     \IF{element[2:$l$] = a}
      \RETURN element
      \STATE \textbf{break}
     \ENDIF
     \ENDFOR
     \ELSE
          \RETURN Map(0a)
      \ENDIF  
  \end{algorithmic}
  \end{algorithm}
Then algorithm \ref{alg:Kernel_map} maps each element of S to an element of A(set of all possible information of length \textit{l}). Let \textit{s} be an element in S, s[i: j] represents the sub-sequence starting from i$^{th}$ position to the j$^{th}$ position of the sequence.\\
(i.e) If \textit{s} = 01101, \textit{s}[2 : 4] = 110. The element \textit{s} $\in$ S is mapped to a $\in$ A if \textit{s}[2 : n-1] = a. 
\begin{example}\label{examp:1}
    Let $n=8$ and $a = 1100101$ then kernel map of $a$ is $111001011$.
\end{example}
For the final part of the encoding, n - 1 redundancy bits are added to each element a of the kernel subset S. Here, i$^{th}$ redundancy bit is obtained using the homomorphism $h_i$. The homomorphism goes as follows
\begin{equation}
h_i(g_1, g_2, \dots, g_n) = \begin{cases} 
      g_{i+1} & 1 \leq i \leq \lfloor{\frac{n-1}{2}}\rfloor\\
      g_1 + g_{i+1} & \lceil{\frac{n+1}{2}}\rceil \leq i\leq n-1 \\
      {g_1 + g_{i+1} + g_{n+1}} & i = \frac{n}{2}, \\
      &\frac{n}{2} \text{ is an integer} 
   \end{cases}
\end{equation}
\begin{example}\label{examp:2}
    The kernel codeword of ‘$a$' from example \ref{examp:1} is 1011010, and the corresponding concatenated codeword $E_a$ = 101101001110.
\end{example}
 \begin{algorithm}[hbt!]
    \caption{Binary code to DNA strings} \label{alg:binarytobp}
    \begin{algorithmic}[1]
      \REQUIRE $E_a$
      \ENSURE encoded information with n bps
      \STATE $n \gets \frac{length(E_a)}{2}$ 
      \FOR{$i = 1 \text{ to } n$}
          \STATE $x[i] \gets$ join($E_a[i], E_a[n+i]$)
        \ENDFOR
        \STATE MAP $00 \Rightarrow C,\text{ } 01 \Rightarrow A$ $10 \Rightarrow T, \text{ } 11 \Rightarrow G$
        \FOR{$ i = 1 \text{ to } n$}
        \STATE  $y[i] \gets $MAP(x[i]) 
        \ENDFOR
        \\\RETURN y
    \end{algorithmic}
  \end{algorithm}
The binary codeword is being separated into small strings of length two with the idea portrayed in algorithm \ref{alg:binarytobp}. This binary to DNA conversion is dependent on an element from each half. In this way, \textit{l} length information is mapped onto the n-length bps. \\
\begin{example}\label{examp:3}
    For $E_a$ = 1110010111101010 from example \ref{examp:2}, the DNA encoded codeword is GGGCATAT.
\end{example} 
This encoding procedure preserves the GC-content of 50\% for even length n, and for odd length, it ranges from 40\% - 60\%. This determines the stability of the code.\\
It gives a Reverse complement distance
\begin{equation}\label{eq:RC}
    d_{rc} = 2 \times \left \lfloor \frac{n-3}{2} \right \rfloor\end{equation}
This construction also has less correlation, and it mostly occurs at the end. This acts as an improvised coding scheme of (Goldman, 2013), where $3^l$ and other $S^l$ methods have been used. Our code achieves the constraint stated in section \ref{sec:Constrains} and serves as a solution to the DNA storage problem.\\
\begin{algorithm}[hbt!]
    \caption{Decoding Kernel codes to binary}
    \label{alg:decode}
    \begin{algorithmic}[1]
       \REQUIRE y \hfill \algorithmiccomment{ DNA codeword }
       \ENSURE information
       \STATE MAP $C, A \Rightarrow 0$ and $T, G \Rightarrow 1$
       \STATE n $\gets$ length(y)
       \FOR{$ i = 2 \text{ to } n$}
       \STATE $\hat{a[i-1]} \gets MAP(y[i])$
       \ENDFOR
       \RETURN $\hat{a}$
    \end{algorithmic}
  \end{algorithm}
The decoding is done using algorithm \ref{alg:decode}, which considers the bases starting from the second position to the last of the codeword and decodes it using the map given. The map is defined from base pair to $\mathcal{Z}_2$.\\
The DNA codeword from example \ref{examp:3} GGCATAT(from the second position) is decoded as follows “T, G” as one and “A, C” as 0. So, it will be 1100101, and the first bits are removed according to the length. Here we assume \textit{l}  = 7, so the information is 1100101.\\
\begin{figure}[ht!]
      \centering
      \includegraphics[width=3in]{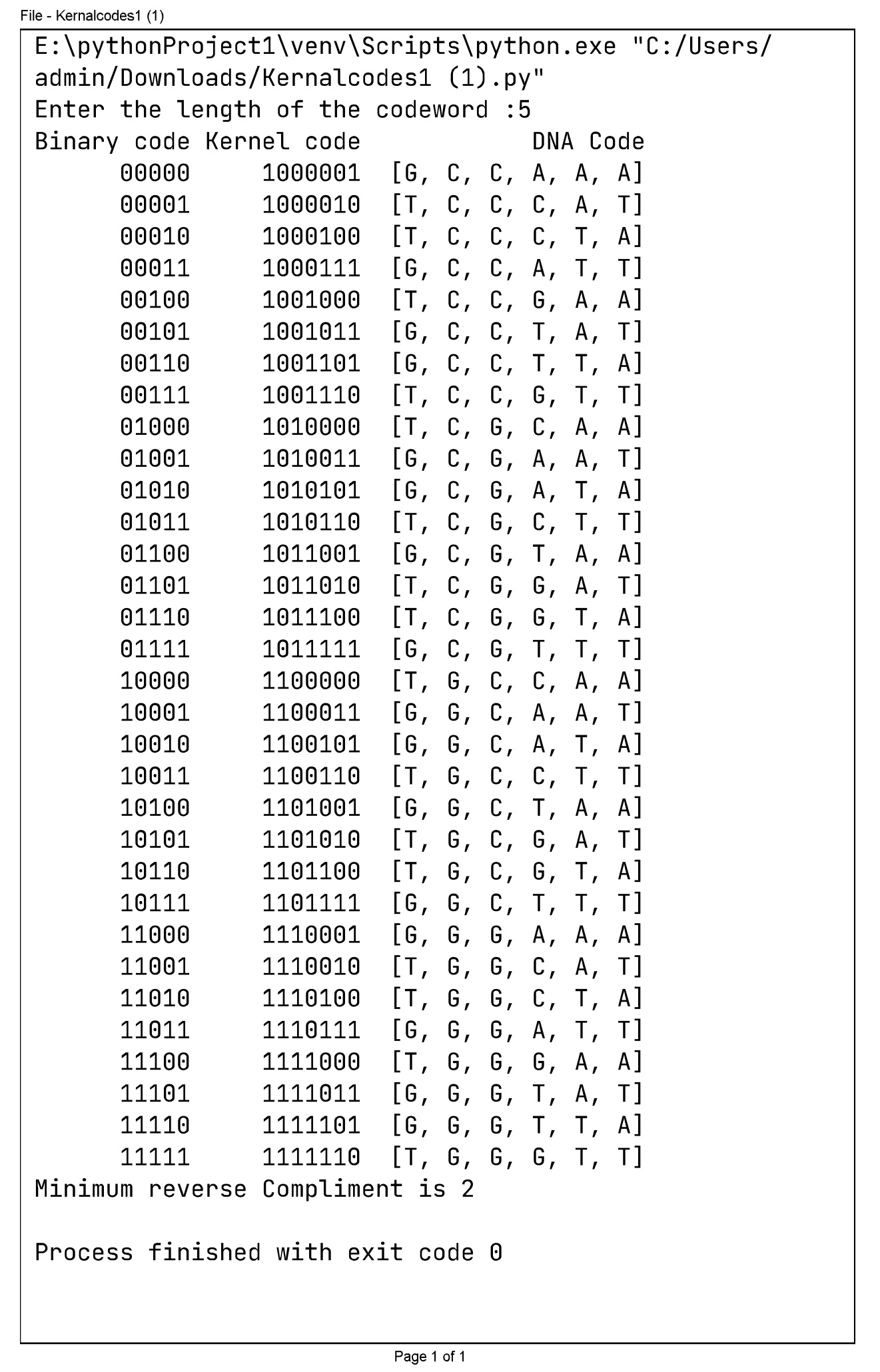}
      \caption{Encoding for five-length information set}
      \label{fig:my_label}
      \noindent\footnotesize{Source: It is the python code output  which was executed on the Pycharm version: 2021.2.2 (open source software) on the computer with the processor Intel(R)Xeon(R) E3-1225 v5@ 3.30GHz.}
  \end{figure}
Figure 1 depicts how the encoding has been done for an information set of length(\textit{l}) = 5 and codeword length(n) = 6 using python3.
\subsection{Comparitive analysis}\label{comparitive analysis}
We compared our code with other existing codes based on the indices, RC length and GC content present in the code. We also examined whether the codeword can be directly mapped from binary information.
\begin{table}[hbt!]
    \centering
    \begin{tabular}{|p{0.2\linewidth} | p{0.3\linewidth}|p{0.15\linewidth} | p{0.15\linewidth}|}
    \hline
         & GC-content & Codes based on proposed RC length & Binary to code without dictionary  \\
         \hline
         Construction of cyclic DNA codes over the ring $Z_4 +vZ_4$ (\cite{bib19})& Less & No & Yes\\
         \hline
         New DNA codes from Cyclic Codes over Mixed Alphabets (\cite{bib11}) & Less & No & Yes\\
         \hline
         Rewritable random access (\cite{bib13})& Maintained only for address with even length & No & No\\
         \hline
         Our paper & Odd: 40\% - 60\%, Even:  50\% & Yes & Yes\\
         \hline
    \end{tabular}
    \caption{Effectiveness of Kernel encoding}
    \label{tab:comparision_others}
\end{table}

The table \ref{tab:comparision_others} demonstrates that our approach is more effective at creating code that complies with RC constraints at any given distance(equation \ref{eq:RC}). Our code can preserve the GC content between 40\% and 60\% for any length. Here we are encoding binary information into the DNA strings, so while retrieving the information, we do not need a separate dictionary to decode it.
  \section{Conclusion}\label{sec13}
   In this paper, the information mapping is done using kernel code, and the proportional GC content is maintained by using concatenated kernel code. Properties of kernel code ensure a certain amount of Reverse complement distance. A comparison of our code with the existing codes reveals that our algorithm increases the reverse complement distance more efficiently. Indel errors are still a place where feature studies can be made.

    \bibliographystyle{apsrev4-1}
    \bibliography{bhavithran_DNA_Paper}
\end{document}